\documentclass[11pt]{article}
\usepackage[intlimits]{amsmath}

\usepackage{charter}

\usepackage{euler}
\usepackage{eucal}

\newcommand{\abs}[1]{\lvert#1\rvert}
\newcommand{\bsigma}{\boldsymbol{\sigma}}

\DeclareMathOperator{\Div}{div}
\DeclareMathOperator{\Rot}{rot}

\allowdisplaybreaks[3]
\begin{document}

\title{On details of the thermodynamical 
derivation of the Ginsburg--Landau equations}

\author{ A.\,V.\,\,Dmitriev$^\dag$\ and W.\,Nolting$^\ddag$
\medskip\\
{\it $^\dag$\ Department of Low Temperature Physics,}\\
{\it The Faculty of Physics, Moscow State University,}\\
{\it Moscow, 119899, Russia.}
\medskip\\
{\it $^\ddag$\ Institut f\"ur Physik,}
{\it Humboldt--Universit\"at zu Berlin,}\\
{\it Invalidenstr.\ 110, Berlin 10115, Germany.}}

\date{\today}
\maketitle

\begin{abstract}

We examine the procedure of thermodynamical derivation of the
Ginsburg--Landau equation for current, which is given unclear
and contradictory interpretations in existing textbooks. We
clarify all steps of this procedure and find as a consequence a
limitation on the validity range of the thermodynamic
Ginsburg--Landau theory, which does not seem to be explicitely
stated up to now: we conclude that the thermodynamic theory is
applicable only to a superconducting specimen that is not a part
of an external current-carrying loop.

\medskip
PACS 74.20.De; Keywords: superconductivity, Ginsburg--Landau
theory.

\end{abstract}

\section{Motivation}

The Ginsburg--Landau theory of superconductivity has been
developed five decades ago and proved to be a powerful and
fruitful method \cite{G1,G2}. However, derivation of its main
equations from the free energy of a superconductor has been
described somewhat briefly in the original paper \cite{GL}, and
some basic points of this procedure seem to remain not completely
understood up to now. Standard textbooks
\cite{deGennes,AAA,LP,Schmidt,Tinkham} give no answer or contain
contradictory answers to questions like these: What is the sense
of the free energy variation with respect to the vector
potential of magnetic field? Why must the result of this
variation be zero within a superconductor? Why should one take
the variation with respect to the vector potential but not with
respect to magnetic induction?  Why should a surface integral
that appears in the process of the variation be omitted?

In this paper we are going to clarify these points and we hope
that the physical assumptions that lie in the grounds of the
thermodynamical Ginsburg--Landau theory will hence become
clearer. Our analysis also shows a restriction on the area
covered by the thermodynamic theory, which does not seem to be
explicitely stated up to now.

The fact that the Ginsburg--Landau equations can be also derived
from the microscopic theory \cite{Gorkov} (see also
\cite{deGennes}) does not reduce importance of the thermodynamic
theory. Indeed, in practical calculations the Ginsburg--Landau 
equations are often used in combination with the
Ginsburg--Landau free energy of the superconductor. This
approach needs, however, the proper understanding of the
interconnections between the equations and the free energy. So
one comes again to the questions formulated above.

\section{Thermodynamics in magnetic field}

Let us first remind the reader the standard thermodynamics of
magnetic substances subject to a magnetic field (see, for
example, \cite{LL:EM}). It is convenient to assume that the
field is produced by a combination of electromagnets connected
to some external sources and fed by them. The magnetic part of
a free energy differential equals to the work $\delta R$ of these
sources necessary to change the field by a small variation
$\delta \mathbf B(\mathbf r)$, provided that the temperature of
the magnetic sample remains constant and mechanical work is
absent. The external sources work on both the field and the
sample itself: both have their energies, which cannot be
separated because the sample and the field interact. So the
sample and the magnetic field constitute our system, and we are
looking for their joint free energy.

It is more convenient to calculate first not $\delta R$ itself
but the opposite quantity, that is, the work of the changing
field on the currents in the magnets or, more precisely, on the
external sources of these currents. The Lorentz force itself
cannot work, because it is always perpendicular to the current:
$$
f_{\text L} = \frac 1c [\mathbf j \times \mathbf B] \perp \mathbf j \,.
$$
However, when a magnetic field is varied, a vortex electric
field appears in agreement with the Maxwell equation
	\begin{gather}
\Rot \mathbf E = - \frac 1c \frac{\partial \mathbf B}{\partial t} \,.
	\label{rotE}
	\end{gather}
This field can work, and its work on the currents performed
during a short time $\delta t$ equals
$$
\delta t \int d^3 r \, \mathbf{jE} \,.
$$

The work of the currents or their sources on
our system~--- the sample and the field~--- has the same
absolute value but the opposite sign:
	\begin{gather}
\delta R = - \delta t \int d^3 r \, \mathbf{jE} \,.
	\label{Work}
	\end{gather}
This formula gives simultaneously the free energy
differential $\delta F$ of our system. Let us transform it using
the Maxwell equation
$$
\mathbf j = \frac c{4\pi} \Rot \mathbf H \,.
$$
Then we have
$$
\delta F = - \delta t \frac c{4\pi} 
\int d^3 r \, \mathbf{E} \Rot \mathbf H \,.
$$
As in \eqref{Work}, the integral here is taken over the volume
of the conductors carrying the current. However, we will assume
that the integration goes over the whole infinite space, which
is more convenient for further calculations. This makes no
difference because $\Rot \mathbf H = 0$ outside the wires of the
electromagnets.

Because
	\begin{gather}
\Div [ \mathbf a \times \mathbf b ] = \mathbf b \Rot \mathbf a  - 
\mathbf a \Rot \mathbf b
	\label{VectorEquality}
	\end{gather}  
one obtains, taking $\mathbf a = \mathbf E$ and $\mathbf b =
\mathbf H$, that
	\begin{gather}
\delta F = \delta t \frac c{4\pi} 
\int d^3 r \, \Div [\mathbf E \times \mathbf H] -
\delta t \frac c{4\pi} 
\int d^3 r \, \mathbf{H} \Rot \mathbf E \,.
	\label{GeneralFVariation}
	\end{gather}
The first integral can be transformed into a surface one over a
boundary of the infinite integration volume, that is, over an
infinite surface. This integral is zero because the fields disappear
at an infinite distance from their sources. What about the
second integral, we eliminate the electric field from it using
Eqn.\ \eqref{rotE} and thus obtain the final formula
	\begin{gather}
\delta F = \delta t \int d^3 r \, \frac 1{4\pi} \mathbf{H} 
\frac{\partial \mathbf B}{\partial t} =
\int d^3 r \, \frac{\mathbf H \delta \mathbf B}{4\pi} \,.
	\label{dFviadB}
	\end{gather}
The domain of integration here is the whole infinite space. 

With the help of \eqref{VectorEquality} this expression can be further
transformed to another form:
	\begin{multline}
\delta F = \frac 1{4\pi} \int d^3r \, \mathbf H \delta \mathbf B =
\frac 1{4\pi} \int d^3r \, \mathbf H \Rot \delta \mathbf A \\
= \frac 1{4\pi} \int d^3 r \, \Div [\mathbf{H} \times \delta \mathbf A] +
\frac 1{4\pi} \int d^3r \, \delta \mathbf A \Rot \mathbf H = 
\frac 1c \int d^3r \, \mathbf j \delta \mathbf A \,.
	\label{dFviadA}
	\end{multline}
Here $\mathbf A$ is the magnetic field vector potential. Both
formulae \eqref{dFviadB} and \eqref{dFviadA} represent the work of
current sources on our system (the sample and the magnetic
field).

One can see from \eqref{dFviadB} and \eqref{dFviadA} that
	\begin{align}
\frac{\delta F}{\delta \mathbf B} & = \frac{\mathbf H}{4\pi} \,, &
\frac{\delta F}{\delta \mathbf A} & = \frac{\mathbf j}{c} \,.
	\label{Fvariations}
	\end{align}
We will see below that the latter equality is important for the
Ginsburg--Landau theory. We remind reader that $\mathbf j$ 
is the density of the
external macroscopic conductivity current
that produces the magnetic
field and stays in the Maxwell equation
$$
\Rot \mathbf H = \frac{4\pi}c \mathbf j \,.
$$
In other words, $\mathbf j$ is the current in the electromagnets.

The formulae obtained in this section are derived from
thermodynamics and hence are always valid under the equilibrium
conditions.

\section{The Ginsburg--Landau free energy and derivation of equations}

As in the Landau theory of second order phase transitions, the
Gins\-burg--Landau free energy of a superconductor appears as an
expansion in terms of the so-called condensate wave function,
$\psi$, that replaces the order parameter for the transition
between superconducting and normal states. In the absence of
magnetic field the free energy has the following simlpe form,
typical for a second order phase transition:
$$
F = F_n + \int d^3r \left[ \frac{\hbar^2}{4m} 
\left| \nabla \psi \right|^2 + a \abs{\psi}^2 +
\frac b2 \abs{\psi}^4 \right] .
$$
Here $F_n$ is the free energy of the sample in normal state. 
The first (gradient) term under the integral describes an energy
increase in a non-uniform state of the superconductor.

If a magnetic field is applied to the sample, the latter
expression is slightly modified:
	\begin{gather}
F = F_n + \int d^3r \left[ \frac{\hbar^2}{4m} \left| \left( \nabla -
\frac{2ie}{\hbar c} \mathbf A \right) \psi \right|^2 + a \abs{\psi}^2 +
\frac b2 \abs{\psi}^4 + \frac{B^2}{8\pi} \right] .
	\label{GLFreeEnergy}
	\end{gather}
The structure of the gradient term here follows from the
gauge invariance principle. The last term under the integral is
a density of the magnetic field energy, which must be added to
the terms describing the superconducting transition itself.

As in the preceeding section, the integral in the
Ginsburg--Landau free energy is taken over the whole infinite
space. Of course, the condensate wave function is zero outside
the superconductor, but the magnetic field energy still exists
in the space around it.

The free energy \eqref{GLFreeEnergy} corresponds to a
non-equilibrium state of the sample that is characterized by
a spatial $\psi$ distribution in a given magnetic field.

To find the equilibrium state of the system one makes use of the
fact that the free energy of a system has a minimum in the
equilibrium state as compared to its value in any
non-equilibrium state, provided that the temperature is fixed
and no work is performed on the system. 

The parameter that describes an internal state of a
superconductor is $\psi$. Hence the equilibrium condition can be
found easily by means of taking the free energy variation with
respect to $\psi$ or its complex conjugate $\psi^*$ at given 
$\mathbf B(\mathbf r)$ and $\mathbf A(\mathbf r)$ space
distributions, and setting the result equal to zero. As the
magnetic field is fixed, no work is performed on the system,
which is evident from \eqref{dFviadB} and \eqref{dFviadA}.

Taking the variation, for example,
with respect to $\psi^*$, one finds
	\begin{multline*}
\delta F = \int_{V_s} d^3r \left[ a \psi \delta \psi^* + 
b \abs{\psi}^2 \psi \delta \psi^*  
{\phantom{\frac{\hbar^2}{4m}}} \right. \\
\left. {} + \frac{\hbar^2}{4m} 
\left( \nabla \psi - \frac{2ie}{\hbar c}
\mathbf A \psi \right) 
\left( \nabla \delta \psi^* + \frac{2ie}{\hbar c} \mathbf A
\delta \psi^* \right) \right] .
	\end{multline*}
As all terms here are proportional to $\psi$, the integral
reduces to the volume of the superconductor, $V_s$. Integrating
the term containing $\nabla \delta \psi^*$ by parts, one obtains
	\begin{multline*}
\int_{V_s} d^3r \left( \nabla \psi - 
\frac{2ie}{\hbar c} \mathbf A \psi \right) \nabla \delta \psi^* + 
\int_{V_s} d^3r \, \left( \nabla \psi - 
\frac{2ie}{\hbar c} \mathbf A \psi \right) 
\frac{2ie}{\hbar c} \mathbf A \delta \psi^* \\
= \oint_{\sigma_s} d\bsigma \, \delta \psi^*
\left( \nabla \psi - 
\frac{2ie}{\hbar c} \mathbf A \psi \right) - 
\int_{V_s} d^3r \, \delta \psi^* \, \nabla 
\left( \nabla \psi - 
\frac{2ie}{\hbar c} \mathbf A \psi \right) \\
\shoveright{ {} + \int_{V_s} d^3r \, 
\frac{2ie}{\hbar c} \mathbf A \delta \psi^* 
\left( \nabla \psi - 
\frac{2ie}{\hbar c} \mathbf A \psi \right) } \\
\shoveleft{ = \oint_{\sigma_s} d\bsigma \, \delta \psi^*
\left( \nabla \psi - 
\frac{2ie}{\hbar c} \mathbf A \psi \right) }\\
{} - \int_{V_s} d^3r \, \delta \psi^* \left[ \nabla^2 \psi -
\frac{2ie}{\hbar c} \nabla ( \mathbf A \psi) -
\frac{2ie}{\hbar c} \mathbf A \nabla \psi +
\left( \frac{2ie}{\hbar c} \mathbf A \right)^2 \psi \right] \\
= \oint_{\sigma_s} d\bsigma \, \delta \psi^*
\left( \nabla \psi - 
\frac{2ie}{\hbar c} \mathbf A \psi \right) -
\int_{V_s} d^3r \, \delta \psi^* 
\left( \nabla - \frac{2ie}{\hbar c} \mathbf A \right)^2 \psi \,,
	\end{multline*}
where $\sigma_s$ is the surface of the superconductor. Now we
substitute the result back into the previous expression:
	\begin{multline*}
\delta F = \int_{V_c} d^3r \left[ a \psi \delta \psi^* + 
b \abs{\psi}^2 \psi \delta \psi^* - 
\frac{\hbar^2}{4m} \delta \psi^* 
\left( \nabla - \frac{2ie}{\hbar c}
\mathbf A \right)^2 \psi \right] \\
{} + \frac{\hbar^2}{4m} \oint_{\sigma_s} d\bsigma \, 
\delta \psi^* \left( \nabla \psi - 
\frac{2ie}{\hbar c} \mathbf A \psi \right) .
	\end{multline*}

There are two contributions to the variation, the bulk term and
the surface one. Taking into account that the superconducting
specimens may have arbitrary sizes and shapes, one can hardly
expect that any compensation of these two terms may ever happen.
So each one must be set equal to zero independently of the
other.

Let us first consider the volume integral. One can suppose
it to be more important for a macroscopic body. The
corresponding variation must be zero, and because $\delta
\psi^*$ in the bulk is an arbitrary function, the following
condition must take place in the equilibrium state to minimize
the volume contribution to the free energy:
	\begin{gather}
-\frac{\hbar^2}{4m} \left(\nabla - \frac{2ie}{\hbar c} \mathbf A 
\right)^2 \psi + a \psi + b \abs{\psi}^2 \psi = 0 \,.
	\label{1stGLEquation}
	\end{gather}
This is the first Ginsburg--Landau equation. As one could see
from the derivation, the equation is an equilibrium condition
for the system with the free energy \eqref{GLFreeEnergy}.

Now let us consider the surface term in the free energy
variation. Generally, it is of less importance, because surface
effects are usually small in macroscopic bodies, that is, in the
macroscopic limit. However, if a media surrounding the
superconductor does not influence electrons in the latter, which
is true for the vacuum or a dielectric, then $\delta \psi^*$ may
take arbitrary values at the surface.  Then one can see that to
eliminate the surface integral, the condensate wave function at
the surface of the semiconductor must satisfy the boundary
condition
$$
\left[ \left( \nabla - \frac{2ie}{\hbar c} \mathbf A \right) \psi
\right]_\perp = 0 \,,
$$
where the subscript stands for the component of the vector
perpendicular to the surface.

Let us note for completeness that the boundary condition at the
boundary of a superconductor and a normal metal was found by de
Gennes \cite{deGennes} (see also \cite{G93}), but its derivation
lies outside the thermodynamic theory we consider in this
article.

Now we have to turn to the next step that Ginsburg and Landau
performed in their original work. They took the free energy
variation with respect to the vector potential $\mathbf A$ and
set the result equal to zero. This action may not be as clear as
the preceding calculations, because in contrast to $\psi$,
$\mathbf A$ is not an internal variable of the superconductor,
which can be varied without any restrictions. The magnetic field
is produced by external electromagnets, it must satisfy the
Maxwell equations, and this restricts the possible vector
potential variations.

As far as we know, no explanation of this point has ever been
given. In some textbooks the words from the original article
were simply reproduced without any comments, whereas other
authors interpreted this procedure as a minimization with
respect to the vector potential.

The latter interpretation seems questionable. First, it is not
clear why one has to minimize the free energy with respect to
the vector potential but not with respect to the magnetic field
itself. These two methods give different results, as is evident
from equalities \eqref{Fvariations}.

Second, as it was already mentioned above, free energy has
minimum in the equilibrium only when the work of external forces
is absent, which implies that $\mathbf B$ space distribution is
fixed, see \eqref{dFviadB}. However, Ginsburg and Landau did not
use this condition. Were it taken into account, it would
completely change their variation procedure.

We think that to understand the actual sense of the free energy
variation with respect to the vector potential, one has to
remember the second of Eqns.\ \eqref{Fvariations} that hold in
the equilibrium state. The equation shows that in the
equilibrium this variation is proportional to the density of the
current that produces the magnetic field. This current flows in
the external electromagnets and in the separated superconductor
sample its density is zero.

Indeed, if one describes the magnetic field in a medium in terms
of two fields $\mathbf B$ and $\mathbf H$, as it is adopted in
the Ginzburg--Landau theory, then magnetic properties of the
medium are characterized by its magnetization which creates the
difference between $\mathbf B$ and $\mathbf H$. The
magnetization is connected with local (or `molecular') currents
in the media, whereas $\mathbf H$ is produced by the
conductivity current that is absent in the sample if the latter
is not connected to the external current-carrying loop.

Hence Eqn.\ \eqref{Fvariations} within the superconductor
takes the form
	\begin{gather}
\frac{\delta F}{\delta \mathbf A} = 0 \,,
	\label{zeroFvariation}
	\end{gather}
and this is just the condition that Ginsburg and Landau use. It
is a general equality that is satisfied in the equilibrium in
any insulated magnetic substance and that is not specific for a
superconductor.

Note that it is based on the fact that the sample is not
connected to the external power supply, so that the sample
carries no external current. Consequently, all current in the
superconducting sample can be treated as a local (`molecular')
one that produces magnetization but does not introduce into
$\Rot \mathbf H$. 

On the contrary, if the superconducting sample would be a part
of the external current loop, then $\mathbf j$ would not be zero
in it, and Eqn.\ \eqref{zeroFvariation} could not be applied to
the sample any more. 

The fact that the thermodynamic Ginsburg--Landau theory is
applicable only to the superconductors that are not included
into an external current loop was not explicitly stated nor in
the original work \cite{GL} nor in later textbooks.

It is worth noting that the existing microscopic derivation of
the Ginsburg--Landau equations \cite{Gorkov,deGennes} implicitly
contains the same limitation as the thermodynamic theory. In
\cite{Gorkov} the initial electron Green's function corresponds
to zero current. In \cite{deGennes} the final basis used in the
process of the equations derivation consists of real electron
wave functions, which also correspond to the absence of 
current.

After this discussion let us continue the derivation of the
second Ginzburg--Landau equation. A variation of the joint free
energy of the superconductor and magnetic field is taken with
respect to $\mathbf A$. Please remember that the volume integral
is taken over the whole infinite space:
	\begin{multline}
\delta F = \int d^3r \, 
\delta \left\{ \frac 1{8\pi} (\Rot \mathbf A)^2 +
\frac{\hbar^2}{4m} \left( \nabla \psi - \frac{2ie}{\hbar c}
\mathbf A \psi \right) 
\left( \nabla \psi^* + \frac{2ie}{\hbar c} \mathbf A \psi^* 
\right) \right\} \\
\shoveleft{ \phantom{\delta F} = 
\int d^3r \, \left\{ \frac{\Rot\mathbf A \Rot \delta \mathbf A}
{4\pi} + \frac{\hbar^2}{4m} \delta \mathbf A 
\left[ - \frac{2ie}{\hbar c} \psi \left( \nabla \psi^* + 
\frac{2ie}{\hbar c} \mathbf A \psi^* \right) \right.\right.} \\
\shoveright{ \left.\left. {} + \frac{2ie}{\hbar c} \psi^*
\left( \nabla \psi - \frac{2ie}{\hbar c} \mathbf A \psi 
\right) \right] \right\} } \\
\shoveleft{ \phantom{\delta F} = 
\int d^3r \, \left\{ \frac{\delta \mathbf A}{4\pi} \Rot \mathbf B -
\frac 1{4\pi} \Div [ \mathbf B \times \delta \mathbf A]  \right. }\\
\left. {} + \delta \mathbf A \left[ \frac{i\hbar e}{2mc} 
\left( \psi^* \nabla \psi - \psi \nabla \psi^* \right) + 
\frac{2e^2}{mc^2} \abs{\psi} \mathbf A \right] \right\} .
	\label{dFbydA}
	\end{multline}
Here again the equality \eqref{VectorEquality}
was used, with $\mathbf a = \Rot \mathbf A$ and $\mathbf b = 
\delta \mathbf A$. 

The integral of the second term in \eqref{dFbydA} is transformed
into a surface integral over an infinite surface and disappears.
This is a consequence of the fact that the integration in the
Ginsburg--Landau free energy is taken over the whole infinite
space, and not over the superconducting specimen only.

Now, setting $\delta F$ equal to zero within the superconductor, 
one obtains the second Ginsburg--Landau equation:
	\begin{gather}
\Rot \mathbf B = \frac{4\pi}c \mathbf j \,, 
	\label{2ndGLEqn-1} \\
\mathbf j = -\frac{i\hbar e}{2mc} \left( \psi^* \nabla \psi - 
\psi \nabla \psi^* \right) -
\frac{2e^2}{mc^2} \abs{\psi} \mathbf A \,. 
	\label{2ndGLEqn-2}
	\end{gather}
As we explained above, it follows from a general equilibrium
equality \eqref{Fvariations}.

Let us also consider, for completeness, the space outside the
superconducting specimen. Here the free energy variation equals
$\mathbf j / c$, as it follows from the second Eqn.\
\eqref{Fvariations}. Because in the space outside the sample
$$
\mathbf B = \mathbf H \,,
$$
one comes in this region to the equation
$$
\Rot \mathbf H = \frac{4\pi}c \mathbf j \,,
$$
where $\mathbf j$ is the external current density in the
electromagnets. This is simply the Maxwell equation, as it
should be.

\section{Conclusion}

So we can conclude that the Ginsburg--Landau equation for
current is derived from the general equality
$$
\frac{\delta F}{\delta \mathbf A} = \frac{\mathbf j}{c}\,,
$$
where $\mathbf j$ is the current density in the electromagnets
that produce the magnetic field. This condition is not specific
to superconductors. It is important also that the integral in
the Ginsburg--Landau free energy is taken over the whole infinite
space so that the magnetic field energy outside the
superconductor is also taken into account.

To obtain the Ginsburg--Landau equations from the superconductor
free energy, one must set $\mathbf j$, the external current
density, equal to zero within the superconductor sample, which
means that the sample is not a part of the external
current-carrying loop. To the best of our knowledge, this
limitation on the validity range of the thermodynamic
Ginsburg--Landau theory has never been explicitly stated.


\end{document}